# Ultrahigh Elastically Compressible and Strain-Engineerable Intermetallic Compounds Under Uniaxial Mechanical Loading

Gyuho Song,[1] Vladislav Borisov,[2] William R. Meier,[3] Mingyu Xu,[3] Keith J. Dusoe,[1] John T. Sypek,[1] Roser Valentí,[2] Paul C. Canfield,[3] Seok-Woo Lee[1]*

**Affiliations**

[1]Department of Materials Science and Engineering & Institute of Materials Science, University of Connecticut, 97 North Eagleville Road, Unit 3136, Storrs CT 06269-3136, USA

[2]Institute of Theoretical Physics, Goethe University, Frankfurt am Main, D-60438 Frankfurt am Main, Germany

[3]Ames Laboratory & Department of Physics and Astronomy, Iowa State University, Ames IA 50011, USA

Electronic mail: seok-woo.lee@uconn.edu

**Abstract:**

Intermetallic compounds possess unique atomic arrangements that often lead to exceptional material properties, but their extreme brittleness usually causes fracture at a limited strain of less than 1% and prevents their practical use. Therefore, it is critical for them to exhibit either plasticity or some form of structural transition to absorb and release a sufficient amount of mechanical energy before failure occurs. This study reports that the $ThCr_2Si_2$-structured intermetallic compound ($CaFe_2As_2$) and a hybrid of its structure ($CaKFe_4As_4$) with 2 μm in diameter and 6 μm in height can exhibit superelasticity with strain up to 17% through a reversible, deformation-induced, lattice collapse, leading to a modulus of resilience orders of magnitude higher than that of most engineering materials. Such superelasticity also can enable strain engineering, which refers to the modification of material properties through elastic strain. Density Functional Theory calculations and cryogenic nanomechanical tests predict that superconductivity in $CaKFe_4As_4$ could be turned on/off through the superelasticity process, before fracture occurs, even under uniaxial compression, which is the favorable switching loading mode in most engineering applications. Our results suggest that other members with the same crystal structure (more than 2500 intermetallic compounds), and substitution series based on them should be examined for the possibility of manifesting similar superelastic and strain-engineerable functional properties.

Intermetallic compounds often exhibit superior/exceptional physical and chemical properties due to their uniquely ordered atomic arrangements[1] but their practical applications have been significantly limited because most of them are extremely brittle and cannot absorb a sufficient amount of mechanical energy before failure occurs.[2] Their rigid covalent/ionic bonds and complex crystal structures usually do not permit plastic deformation or structural transition, leading to brittle failure at an elastic limit of less than 1%. Therefore, it is extremely rare to obtain a large elastic limit over 10% in intermetallic compounds except for some special cases such as shape memory alloys where the heat or magnetic field induces the large strain recovery.[3-8]

Recent studies on mechanical behavior of materials at the nano-/micro-meter scale revealed that a material could sustain a higher stress and higher elastic limit as its dimension decreases.[9-15] Particularly for a brittle material, according to the weakest link mechanism, a smaller sample contains a smaller number of defects statistically, leading to a higher fracture strength as well as a higher fracture strain.[16-18] For instance, nanowires and nanoparticles often exhibit ultrahigh elastic strain, compared to their corresponding bulk materials.[19-21] Large elastic deformation at the nano-/micro-scale could induce substantial changes in structure and material property and could enable strain engineering, which refers to the modification of material properties through elastic strain.[15] One of the most known examples would be the increase in carrier mobility in elastically-strained silicon under bi-axial strain.[22] Because the size effect on elastic strain has been widely observed in brittle materials at small length scales, it could be seen also in novel intermetallic compounds at the nano-/micro-meter scale.

Recently, high temperature Fe-based superconductors have drawn strong attention due to their superconducting capability even in the presence of magnetic Fe, which has been regarded as a harmful element for superconductivity.[23-24] Now, they are regarded as a great material system that allows the study of relationship between superconductivity and magnetism. ThCr$_2$Si$_2$-structured Fe-based pnictides have been extensively studied due to their strong pressure sensitivity of structure and electronic/magnetic properties.[25] Particularly, CaFe$_2$As$_2$ single crystals undergoes the collapsed tetragonal (cT) phase transition, which leads to ~10% reduction of c-axis lattice parameter under hydrostatic pressure.[26] We also performed uniaxial mechanical tests on CaFe$_2$As$_2$ micropillars and observed unique mechanical behaviors including ~13% of superelasticity, superior fatigue resistance, and cryogenic shape memory effects, and micaceous plasticity.[27, 28] Notably, magnetism of CaFe$_2$As$_2$ changes from paramagnetic (or antiferromagnetic) to nonmagnetic states when the cT transition occurs.[26] Thus, strain engineering of magnetism is possible for CaFe$_2$As$_2$ or its related structures.

Recently, the hybrid structures of Fe-based pnictides, CaKFe$_4$As$_4$, has been actively investigated due to its high temperature superconductivity (T$_c$~35 K).[29] The previous study confirmed that superconductivity can be switched off under hydrostatic pressure through the half collapsed tetragonal transition (hcT) around Ca atom.[29] However, uniaxial mechanical tests on CaKFe$_4$As$_4$ have never done, and it should be interesting to see how differently CaKFe$_4$As$_4$ behaves, compared to CaFe$_2$As$_2$ in terms of superelasticity. The insertion of large K atoms into the lattice makes the As-As distance around K atom larger. The larger interplanar spacing could allow the larger linear elastic strain simply because the widely-spaced layers could be compressed more. However, the large As-As distance would make the As-As bond formation more difficult. Thus, the competition of these two factors could affect the total elastic strain. In addition, we noticed that the plastic slip (or shear fracture) of CaFe$_2$As$_2$ occurs in the 1/3[3 1 $\bar{1}$](1 0 3) slip system under compression along c-axis. Due to the larger c-axis length of CaKFe$_4$As$_4$,[29] the slip vector of CaKFe$_4$As$_4$ in the same slip system should be larger than that of CaFe$_2$As$_2$, implying that CaKFe$_4$As$_4$ would exhibit the higher yield strength, based on the Peierls-Nabarrow model.[30, 31] In this sense, $CaKFe_4As_4$ could exhibit the elastic strain larger than that of $CaFe_2As_2$.

In this study, therefore, we performed uniaxial micropillar compression tests and Density Functional Theory calculation to investigate the superelastic properties of $CaKFe_4As_4$ and compared the mechanical data with those of $CaFe_2As_2$. We found a giant elastic limit, up to 17%, in $CaKFe_4As_4$, under uniaxial compression along the c-axis of their unit cells. Density functional theory calculations revealed that its enormously large elastic strain primarily results from atomic bond formation around Ca atom and local elastic compliance around K atom. Also, the cyclic compression test on $CaKFe_4As_4$ showed that the superelastic deformation over 10% strain is completely reversible when the applied force is relaxed. This uniaxial process is entirely distinct from the conventional shear-based superelastic mechanism, martensite-austenite phase transformation of shape memory alloys and ceramics.[3-8] Moreover, we also used our custom-built in-situ cryogenic micromechanical testing system to investigate the effects of temperature on the first hcT transition, which is known to remove superconductivity in $CaKFe_4As_4$. We found that the onset stress of the first hcT transition near the superconducting transition temperature is much lower than the fracture strength. This result suggests that there is a strong possibility to see the superconductivity switching even under uniaxial compression before fracture occurs.

$CaFe_2As_2$ is grown from a Sn-rich solution, and $CaKFe_4As_4$ was grown from excess FeAs. The constituent materials were put in alumina crucibles, which are located in an amorphous silica ampule. Single crystals were slowly grown under slow cooling in a furnace and were quickly decanted using a centrifuge. The detailed descriptions of the solution growth of our crystals are also available elsewhere.[32, 33] Micropilllars are fabricated using focused-ion beam milling, Helios Nanolab 460F1 (Thermo Fisher, USA). Gallium ion beam currents from 300 to 10 pA under an operating voltage of 30 kV were used from the initial to final thinning with concentric circle patterns. Because the typical thickness of FIB damage layer is about 20 nm, which is much thinner than our pillar diameter (~2 μm), we expect negligible effects of FIB damage on the mechanical data. *In-situ* nanomechanical test was performed at room temperature and under an ultra-high vacuum condition (<10$^{-4}$ Pa) using a NanoFlip$^{TM}$ (KTL-Tencor, USA), which is installed in a field-emission gun JEOL 6335F scanning electron microscope (JEOL, Japan). A nominal displacement rate of 10 nm/s, which corresponds to the engineering strain rate of ~0.002 s$^{-1}$, was used for all *in-situ* compression tests in this study. Strain calculations for experiments were done with the Sneddon punch correction using the effective Young's modulus that can be measured by nanoindentation.[34] The recorded video was often used to visually confirm that our strain measurements were accurate. A liquid nitrogen and helium cryostat, ST-100, was used to perform nanomechanical testing at low temperatures to investigate the temperature effects on mechanical properties. The temperature of the diamond tip was maintained to be similar with that of the sample by using thermal equilibration, leading to a thermal drift below 0.5 nm/sec at all times. A detailed description of our cryogenic system is also available in the Supplementary Information (Supplementary Figure S1 and S2, and Supplementary Note 1 in the supplementary material). Contact stiffness was measured during compression testing by applying a force oscillation with 10 nN in amplitude and 200 Hz in frequency and measuring the resultant displacement oscillation. Contact stiffness data often shows the clearer evidence of lattice collapse than stress-strain curve.[35] The contact stiffness, $S$, is measured by

$$S = \left[ \frac{1}{\frac{P_{OS}}{h(\omega)}\cos\phi - (K_S - m\omega^2)} - K_f^{-1} \right]^{-1} \quad (1)$$

, where $P_{OS}$ is the magnitude of force oscillation, $h(\omega)$ the magnitude of resulting displacement oscillation, $\omega$ is the frequency of oscillation, $\phi$ the phase angle between the force and displacement signals. $K_S$ and $K_f$ are the stiffnesses of the leaf spring and indenter frame, respectively.[34] When structural collapse occurs, the phase angle becomes smaller, providing a lower contact stiffness. This method is useful to capture the availability of structural collapse during micromechanical test.

For Density Functional Theory calculations, structure optimization in the CaKFe$_4$As$_4$ system was performed using state-of-the-art projector-augmented wave method[36] and the generalized-gradient approximation[37] available in VASP code.[38] The energy cutoff was set to 800 eV and the k-mesh dimensions were (5x5x5). Although the system doesn't show a long-range magnetic order, the inclusion of Fe local moments in the simulation is necessary for a correct description of structural transitions under pressure, as detailed in previous studies.[29,39] For that reason, we imposed the "frozen" spin-vortex spin configuration on the Fe sublattice which approximates the effect of spin fluctuations present in this material.[39] In the current work, we simulate the uniaxial [001]-strain conditions by varying the c-lattice parameter and calculating the total energies of CaKFe$_4$As$_4$ structures optimized for different a-lattice parameters. Fit to the Birch-Murnaghan equation of state

$$E(V) = E_0 + \frac{bV}{b_0(b_0-1)}\left[b_0\left(1 - \frac{V_0}{V}\right) + \left(\frac{V_0}{V}\right)^{b_0} - 1\right] \qquad (2)$$

allowed to estimate the equilibrium lattice parameters for a given value of strain. In the final step, the internal atomic positions are optimized for the fixed lattice dimensions and the stress value along the c-axis was obtained. The electronic properties of the optimized structures were calculated using the all-electron full-potential localized orbitals basis set (FPLO) code[40] within the GGA approach. The half-collapsed tetragonal transition was captured by inspecting the energy position of the As 4p$_z$ antibonding orbitals near the Ca layer.[29,39] In order to illustrate the two half-collapse transitions in CaKFe$_4$As$_4$ under the uniaxial load, we plot the real-space distribution of the electron density associated to the As 4p$_z$ orbitals across both Ca and K layers at different pressures. These density maps were obtained from the Wannier functions calculated using the FPLO code[40] and the tricubic interpolation[41,42] on a three-dimensional grid.

Bulk single crystals (Figs. 1(a) and 1(b)) of ThCr$_2$Si$_2$-type intermetallic compound (CaFe$_2$As$_2$) and its hybrid structure (CaKFe$_4$As$_4$) were grown using a solution growth method[17, 18], and cylindrical micropillars with ~2 μm in diameter and ~6 μm in height were fabricated along the [0 0 1] direction using focused-ion beam (FIB) milling. Note that the CaKFe$_4$As$_4$ structure can be thought of as a periodic replacement of half of the Ca in CaFe$_2$As$_2$ with K in an alternating order along the c-axis, and it looks like a hybrid of CaFe$_2$As$_2$ and KFe$_2$As$_2$.[29,43] The representative stress-strain data of CaFe$_2$As$_2$ and CaKFe$_4$As$_4$ show a large compressive elastic limit of 10.5 and 17%, respectively (Figs. 1(c) and 1(d)). For comparison, the superelastic regime of the stress-strain data for superelastic zirconia[44] and NiTi[45] micropillars, both of which are well-known superelastic materials, are plotted together with our CaKFe$_4$As$_4$ data (Fig. 1(d)). It is clearly seen that our materials exhibit much greater performance in terms of both yield strength and elastic limit. We identify three stages in the stress-strain data. Interestingly, the non-linear stress-strain responses of our crystals resemble that of typical superelastic shape memory alloys, suggesting that the uniaxial deformation of our crystals would induce a structural transition, too. Also, the deformation is completely reversible when the applied load is relaxed (Fig. S3 in the supplementary material and the inset of Fig. 5 shown below) and is repeatable under cyclic deformation (Ref. 28 and Fig. S4 in the supplementary material). Note that we have never seen any evidences of shear deformation from more than 20 samples tested. The real-time SEM videos always showed a clean surface until fracture occurs. Thus, our superelasticity does not appear to be related to any conventional shear-based mechanism such as martensite-austenite phase transformation, which forms shear bands and causes a significant lateral displacement particularly in the case of a single crystal.[44]

The decrease in contact stiffness within Stage II looks counter-intuitive (Figs. 1(e) and 1(f)) because the contact stiffness of a solid material usually increases during compression.[10] Hoffmann and Zheng hypothesized that such a decrease would be possible through a process of forming and breaking Si-Si-type bonds in $ThCr_2Si_2$-type structures under uniaxial compression along the c-axis.[46] Our previous works on $CaFe_2As_2$ confirmed that the main mechanism of phase transition is the structural collapse through the formation of As-As bonds under uniaxial compression along the c-axis.[27,47-49] Thus, the decrease in contact stiffness indeed results from a strain-induced structural collapse through As-As bond formation in these structures. On the course of deformation, the formation of As-As bonds makes materials more elastically compliant, but once the lattice collapse is almost complete, the contact stiffness increases again. Thus, stage I and III would correspond to the elastic deformation before and after formation of As-As bonds, respectively. Stage II would correspond to the deformation on the course of As-As bond formation.

Note that the elastic limit 15~17% of $CaKFe_4As_4$ is truly extraordinary, compared to any other shape memory intermetallic compounds, as well as $CaFe_2As_2$ (11~14%) in this study (Figs. 1(d) and 2(a)). Our Density Functional Theory (DFT) calculations of $CaKFe_4As_4$ under uniaxial strain find two half-collapsed tetragonal (hcT) transitions for this system (Figs. 2(b), 2(c), and 2(d)), while the full-collapsed tetragonal (cT) transition is observed in $CaFe_2As_2$.[27,48] This result is qualitatively similar to results of application of hydrostatic pressure.[26,29,49,50] Non-spin-polarized electron density associated with the As-$4p_z$ orbitals clearly shows the presence of two separate hcT transitions in $CaKFe_4As_4$ at different strains (Figs. 2(c) and 2(d)). In $CaKFe_4As_4$, the first hcT appears at a strain of ~0.05-0.08 when As atoms around Ca form As-As bonds. The smaller atomic radius (231 pm) of the Ca atom allows a shorter As-As distance (3.107 Å) around it, leading to the formation of As-As bonds under a low compressive strain (~0.05). Thus, the structural transition in Stage II of CaKFe$_4$As$_4$ (Fig. 1(d)) would be related to the deformation after the onset of the first hcT transition. Based on our DFT data, the second hcT occurs at strain values near the experimentally measured elastic limit (fracture strain), ~0.18-0.19. At this transition, As atoms around K form As-As bonds. Since the atomic radius of K (280 pm) creates a longer-distance between the As-As layer (4.205 Å), larger elastic strain is needed to reach the second hcT transition. Thus, Stage III of CaKFe$_4$As$_4$ would correspond to the gradual formation of As-As bonds around K atom. All these results also explain a higher rate of structural collapse per strain for the full collapse in CaFe$_2$As$_2$ (~15 GPa) than that for the half collapse in CaKFe$_4$As$_4$ (~31 GPa) for Stage II deformation (Figs. 1(c) and 1(d)).

We also carefully monitored our DFT data to examine the contribution of As-As layers near a K atom to the total elastic strain. The interlayer distance of As-As layer around a K atom is 3.2816Å near the elastic limit. By considering that it is 4.205Å before compression, its change contributes to (4.205Å-3.2816Å)/(12.6205 Å, initial c-length) ≈ 0.073 of strain (~41% of the total elastic limit), which is remarkably high. (the third figure in Fig. 2(d) and Fig. S5 in the supplementary material). This result implies that the larger atom size of K makes the formation of As-As bond more difficult but makes the region between As-As layers around the K atom more elastically compliant. Also, note that our elastic limit, ~17%, is close to the elastic strain at which the second hcT occurs in our DFT data. The driving force of the second hcT, i.e., the formation of As-As bonds around the K atom would partially contribute to a large elastic compliance. In view of these observations, we attribute the extraordinary elastic limit of CaKFe$_4$As$_4$ to the presence of these two hcT transitions and the larger atomic size of K. Therefore, it is important to control bond formation and local elastic compliance to tune the superelastic properties.

The elastic limits of our intermetallic compounds are exceptionally high, compared to other superelastic materials even in similar length scales (Fig. 3(a), Fig. S6 and Supplementary Note 2 in the supplementary material). The comparable superelastic strain can be observed only when the dimension of shape memory intermetallic compounds become close to 100 nm. Usually, materials become stronger when the sample dimension is reduced to the nanometer length scale for various reasons.[9-19] For instance, NiTi nanopillars with 150 nm in diameter exhibit an improved yield strength, leading to a 15% elastic limit.[51] If the micrometer length-scale of our specimen is considered, the observed elastic limit (10~17%) is absolutely outstanding. Note that we also observed size effect in $CaFe_2As_2$. Its sub-micron sized pillars exhibit ~17% elastic strain (Fig. S7 in the supplementary material). This size effect could be related to the weakest-link mechanism.[16] Brittle materials often exhibit higher strength and higher elastic limit when they become extremely small. This has been observed in various brittle materials, such as ceramic, diamond, metallic glass, and nanowires, and the weakest-link mechanism is one of the most widely accepted ideas.[16-18,52] If a sample dimension becomes larger, it is more likely to have weaker defects that can induce brittle failure at a lower strength. Vice versa, fracture strength (also, elastic limit) increases as the sample dimension becomes smaller because it is unlikely to find the weak defect. If the diameter of our specimens is also reduced further down to sub-100 nm, it would be possible for them to show much greater performance, and the detailed analysis on the size effect will presented in a separate work.

It is also worthwhile to compare the elastic performance with other advanced engineering materials. The Ashby Chart has been extensively used when material properties of new materials need to be compared with those of other materials.[53-55] Our materials are located in the $E$-$\sigma_y$ space of the Ashby chart (Fig. 3(b)), where $\sigma_y$ is yield strength and $E$ is Young's modulus.[56] Due to the non-linearity of the stress-strain curve, the effective Young's modulus ($E_{eff}$) can be estimated by $E_{eff} = \frac{\sigma_y^2}{2R}$, where $R$ is the modulus of resilience. The modulus of resilience is the maximum mechanical energy absorption per unit volume prior to yielding and can be calculated by integrating the stress-strain curve from 0 to the elastic limit. In the $E$-$\sigma_y$ space of the Ashby chart, as a material is located closer to the right-bottom corner, it can absorb higher mechanical energy per unit volume. Note that both $CaFe_2As_2$ and $CaKFe_4As_4$ are located in the white space, indicating their superior elastic performance to absorb large amounts of strain energy before yielding (Fig. 3(b)). The average moduli of resilience are 143 MJ/m$^3$ and 291 MJ/m$^3$ for $CaFe_2As_2$ and $CaKFe_4As_4$, respectively (Figs. S8 and S9 in the supplementary material). By considering the accurate measurement of stress-strain curve and their reproducibility, our $R$ data are accurate for our micropillar samples. Note that the total strain energy absorption of our specimen are orders of magnitude higher than most engineering materials at both bulk and micrometer scales such as elastomers (~1 MJ/m$^3$), advanced composites (~0.5 MJ/m$^3$), conventional shape memory alloys (~50 MJ/m$^3$), metallic nanopillars (~10 MJ/m$^3$), and superelastic ceramic micropillars (~50 MJ/m$^3$). Semiconductor or ceramic nanowires sometimes show the extremely high modulus of resilience (> 1000 MJ/m$^3$) due to their defect-free structure in their ultra-thin diameter (~50 nm) (Fig. 3(b) and Supplementary Note 3). In sum, Figs. 3(a) and (b) shows the giant superelasticity of $CaKFe_4As_4$ as well as the great potential of Fe-based pnictide superconductors as a superelastic material.

The giant elastic limit of our materials could make strain-engineering possible. Strain-engineering refers to a significant modification of the properties of solid materials by applying an elastic strain.[15] As a similar material with ours, Co-doped $CaFe_2As_2$, $Ca(Fe_{1-x}Co_x)As_2$, shows the superconductivity switching by application of bi-axial deformation on *ab* plane, which changes the *c*/*a* ratio of the unit cell and results in the significant shift of superconductivity region in the temperature-composition phase diagram.[57,58] This strong effect of the *c*/*a* ratio on superconductivity could extend to our $CaKFe_4As_4$ system, which is also a high temperature superconductor ($T_c \approx 35K$), under c-axis uniaxial compression that should change the *c/a* ratio much more significantly.

Previous experimental study with hydrostatic pressure demonstrated that superconductivity of CaKFe$_4$As$_4$ can be turned off reversibly by the application of hydrostatic pressure of 4GPa when the system undergoes the first hcT transition.[29] That is, the formation of As-As bonds around the Ca atoms is a key process to turn off superconductivity. We recently developed an *in-situ* cryogenic micro-mechanical testing system (Figs. S1 and S2, and Supplementary Note 1 in the supplementary material) and confirmed that the first hcT transition occurs only around 1 GPa under uniaxial compression without failure (Fig. 4(a)). Interestingly, the structural collapse behavior in CaKFe$_4$As$_4$ is insensitive to a change in temperature, whereas pure and Co-doped CaFe$_2$As$_2$ exhibits strong temperature sensitivity (Fig. S10 in the supplementary material). These different temperature dependences under uniaxial stress are consistent with the hydrostatic pressure experimental data.[29,49]

The weak temperature dependence of CaKFe$_4$As$_4$ is not fully understood, yet. However, there are several indirect experimental and computational evidences to explain the different temperature sensitivity between CaKFe$_4$As$_4$ and CaFe$_2$As$_2$. Several computational studies suggested that the length between atomic layers in Fe-based pnictides are strongly affected by the distribution of magnetic moments.[59] Interestingly, the magnetic susceptibility of CaKFe$_4$As$_4$ exhibit the weak temperature dependence.[60] This result implies that the magnetic structure or the spin ordering does not change with temperature, so that the c-axis does not change much with temperature.[60] As a consequence, mechanical properties are insensitive to temperature, too. This scenario is consistent with our experimental data in Fig. 4(a).

CaFe$_2$As$_2$ behaves differently. Magnetic susceptibility decreases with temperature when it is the paramagnetic tetragonal structure.[32] Inelastic neutron scattering measurement showed that as the temperature decreases, the short-range antiferromagnetic ordering increases.[61] At the same time, the c-axis length decreases substantially with temperature.[62] Thus, magnetism and structure of $CaFe_2As_2$ are more sensitive to temperature than $CaKFe_4As_4$. Interestingly, once $CaFe_2As_2$ becomes antiferromagnetic orthorhombic structure at a temperature below the transition temperature, it becomes magnetically rigid because magnetic ordering does not occur anymore. Then, the c-axis length of $CaFe_2As_2$ does not change much with temperature,[62] as that of $CaKFe_4As_4$ does. This is probably why the onset stress of cT transition drastically decrease when $CaFe_2As_2$ is tetragonal but becomes nearly constant once $CaFe_2As_2$ becomes antiferromagnetic orthorhombic structure under hydrostatic pressure[49] and uniaxial stress (Supplementary Figure S10c shows the weak temperature dependence between 40 K and 100 K.).

Our DFT simulation confirms that the formation of As-As bonds and the change in electronic structure under uniaxial compression are nearly similar to those under hydrostatic compression (Figs. 4(b) and 4(c)). A similar situation has been found in $CaFe_2As_2$ when comparing the electronic structure under uniaxial and hydrostatic stress.[48] These results make sense because the deformation along c-axis is much more significant than that along a- and b-axes direction even under hydrostatic pressure due to the large elastic compliance and As-As bond formation along c-axis. Under the assumption that the origin of superconductivity is linked to the electronic and magnetic properties of the system, these results suggest that superconductivity could also be switched off by inducing the first hcT transition (the formation of As-As bond around a Ca atom) under uniaxial compression. Therefore, our experimental and computational results strongly suggest that superconductivity of $CaKFe_4As_4$ could be reliably turned on and off at the onset of the first hcT transition (~1 GPa) without any fracture or plastic deformation, even under uniaxial compression. By considering the presence of the onset and offset of the first hcT transition in Stage II, $CaKFe_4As_4$ would begin to show the finite resistivity at the uniaxial stress above ~1GPa and would become completely non-superconducting above ~2.7GPa (Fig. 5). Note that our result does not provide the direct evidence of superconductivity switching, yet. Resistivity measurement or permanent magnetic field measurement would be necessary under uniaxial compression to prove our prediction. By considering all experimental and computational data, superconductivity switching is likely to occur even under uniaxial deformation. Thus, electrical and magnetic measurements at the small length scales are considered as the next step we would like to pursue.

Strain-engineering is usually possible when a material can absorb a large amount of strain without permanent deformation. Hydrostatic stress or bi-axial stress would often be regarded as a convenient way to see the strain effect because the maximum shear stress is zero or too low to cause plastic deformation or fracture. Thus, it is really rare to see substantial strain effect on material properties in brittle intermetallic compounds, particularly under a uniaxial strain condition, because the shear stress is usually sufficiently high to cause brittle failure too easily. In contrast, the structural transition in our intermetallic compounds, $CaFe_2As_2$ and $CaKFe_4As_4$, through formation of covalent bonds leads to a giant uniaxial elastic strain, which can cause substantial changes in their electronic and magnetic properties before failure, even under uniaxial mechanical loading. The uniaxially-loaded superconductivity switching capability may never be imagined in conventional oxide-base superconductors, which have no superelasticity mechanism and easily shatter at a small elastic limit under uniaxial stress due to their extreme brittleness. In addition, our previous experimental studies also demonstrated that $CaFe_2As_2$ could exhibit shape memory effect and thermal actuation under cryogenic environments and have a strong potential for cryogenic actuation technology for space exploration.[27] Some groups of $CaKFe_4As_4$ structured intermetallic compounds are regarded as quantum materials that exhibit unique electronic and magnetic properties.[50] More interestingly, $ThCr_2Si_2$-type and its related structures have been considered to be one of the most populous of all crystal structure types.[63] There are nearly 2500 $ThCr_2Si_2$-structured intermetallic compound.[25] If we consider their hybrid structure, there could be a much larger number of similar intermetallic compounds.[39,43] Also, their microstructure and composition can easily be tuned through heat treatment and solid solutionization.[64] Thus, our observation can be extended to search for a large group of superelastic and strain-engineerable functional materials. Computer simulations with machine-learning could be extremely beneficial to rapidly identify compounds with these desired properties.[65,66] In sum, our discovery of superelasticity and strain-engineerability under "uniaxial" mechanical loading will lead to a grand research opportunity in materials science, solid-state physics, device engineering, and computer simulations.

## SUPPLEMENTARY MATERIALS

See supplementary material for the additional mechanical data, the description of in-situ cryogenic nanomechanical testing system, and in-situ videos.

## ACKNOWLEDGMENTS

G. Song, K.J. Dusoe, J.T. Sypek, and S.-W Lee acknowledge support from UConn Research Excellence Program Funding, the Early Career Faculty Grant from NASA's Space Technology Research Grants Program, and GE Fellowship. FIB work was performed using the facilities in the Uconn/Thermo Fischer Scientific Center for Advanced Microscopy and Materials Analysis (CAMMA). Work by P.C. Canfield, W.M. Meier, and M. Xu was supported by the U.S. Department of Energy, Office of Basic Energy Science, Division of Materials Sciences and Engineering. Their research was performed at the Ames Laboratory. Ames Laboratory is operated for the U.S. Department of Energy by Iowa State University under Contract No. DE-AC02-07CH11358. W.M. Meier was also funded by the Gordon and Betty Moore Foundation's EPiQS Initiative through Grant GBMF4411. Work by R. Valentí and V. Borisov was supported by DFG Sonderforschungsbereich TRR 49, the centre for supercomputing (CSC) in Frankfurt.

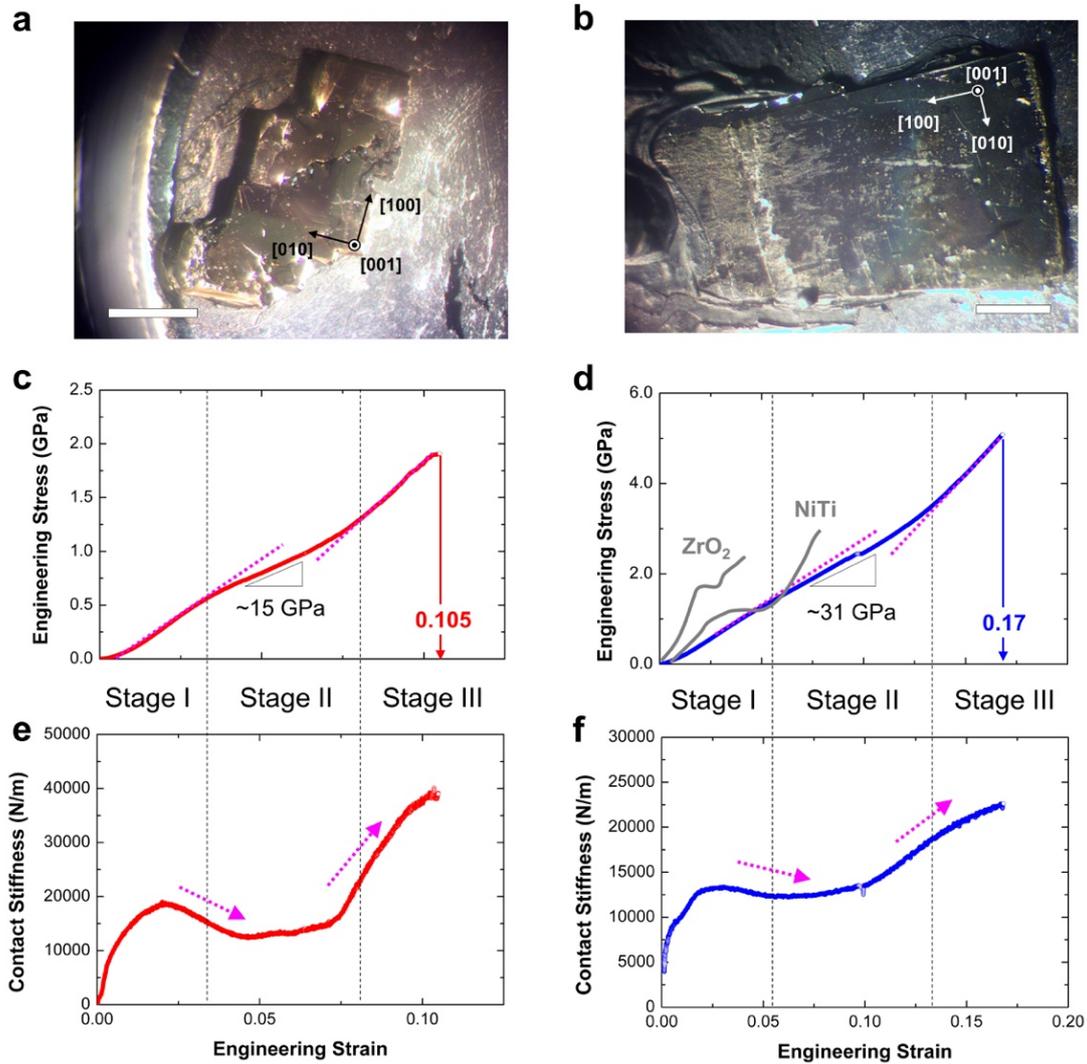

**FIG. 1 Solution-grown single-crystalline intermetallic compounds and room temperature mechanical data.** Optical micrographs of (a) $CaFe_2As_2$ (scale bar, 1mm) and (b) $CaKFe_4As_4$ (scale bar, 1mm); Uniaxial engineering stress-strain data until failure occurs of (c) $CaFe_2As_2$ and (d) $CaKFe_4As_4$; Contact stiffness as a function of engineering strain of (e) $CaFe_2As_2$ and (f) $CaKFe_4As_4$. All stress-strain data exhibit three stages of elastic deformation, which is similar with that of shape memory alloys. The decrease in contact stiffness implies that a material becomes more elastically compliant under compression and corresponds to the structural transition from tetragonal to fully-collapsed ($CaFe_2As_2$) or one or both half-collapsed ($CaKFe_4As_4$) tetragonal structures. For comparison, the stress-strain curves of the elastic regime of superelastic zirconia[44] and NiTi[45] micropillars were added in Fig. 1(d).

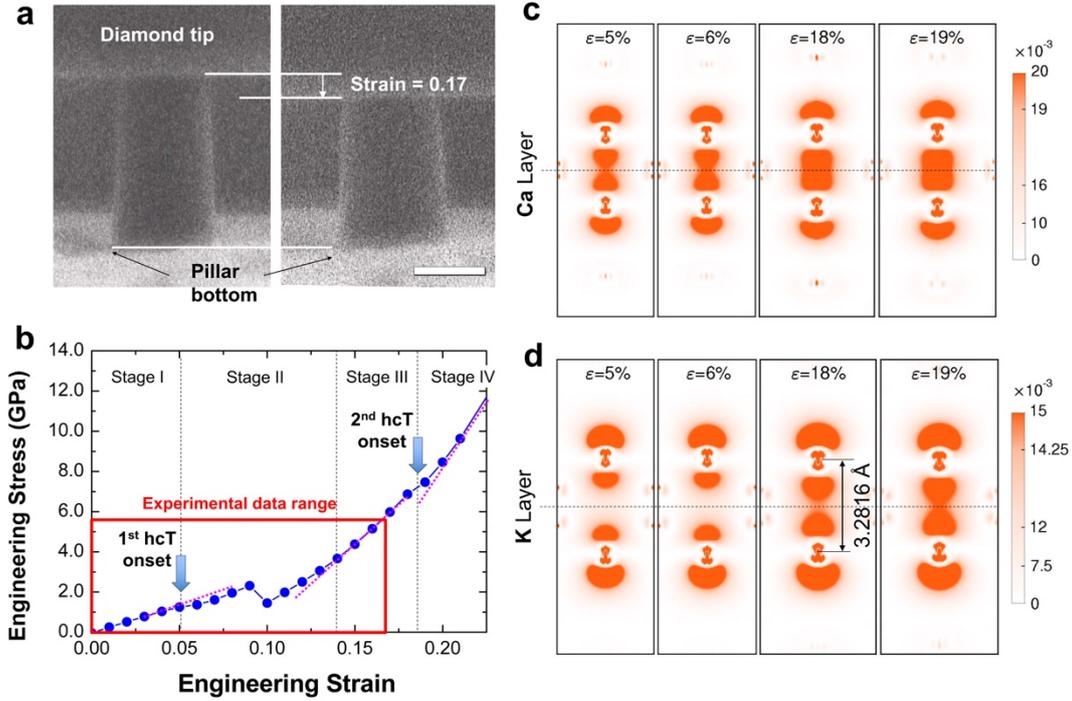

**FIG. 2. Superelasticity of CaKFe$_4$As$_4$.** (a) Snapshots of *in-situ* video right before contact with the diamond tip and right before failure (scale bar, 1μm); (b) DFT simulation results of engineering stress-strain data. Red-line box represents the experimental data range that is limited by fracture. Note that the sharp drop of engineering stress around 0.1 strain occurs due to the collapse of magnetic moments, which are intentionally introduced to mimic paramagnetic state. Due to random distribution of magnetic moments at a finite temperature in a real system, this effect would spread within Stage II. Stage IV corresponds to the elastic deformation after the second hcT transition, which cannot be seen in a real system due to fracture in Stage III.; Non-spin-polarized electron density in the *ac* plane associated with the As-4p$_z$ orbitals near (c) Ca and (d) K at different strains. (c) shows clear bond formation across the Ca-layer by 0.05 strain and (d) shows clear bond formation across K-layer by 0.18 strain.

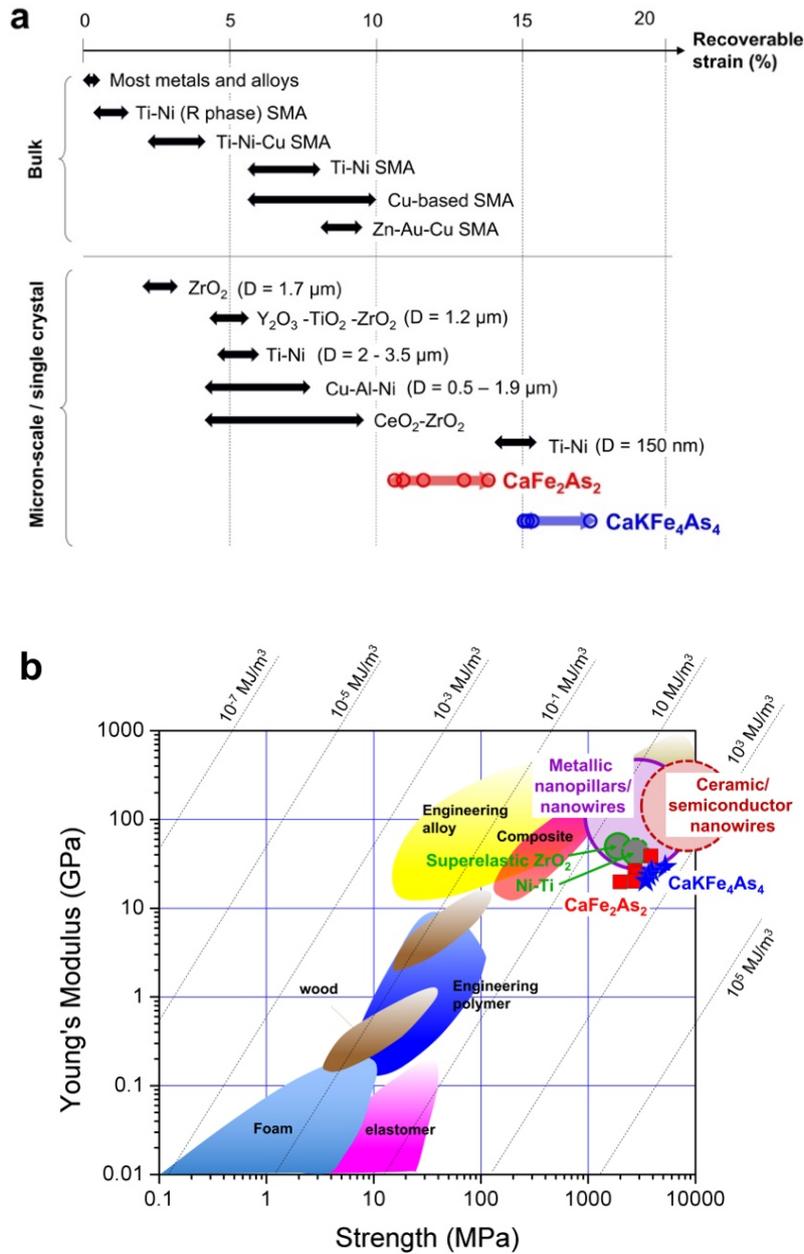

FIG. 3. **Superelastic performance of $CaFe_2As_2$ and $CaKFe_4As_4$.** (a) Elastic limit of superelastic materials at different length scales. The range of recoverable strain (the range of double-headed arrow) shows the minimum and maximum values of our experimental data. Circles in the arrows of our samples indicate the data we obtained (Figs. S8 and S9 in the supplementary material); (b) Ashby Chart of Young's modulus and yield strength. The dotted lines are the contours of modulus of resilience, which is the total strain energy absorption per unit volume prior to yielding. Data of nanopillars and nanowires are available in Supplementary Information (See Supplementary Note 3).

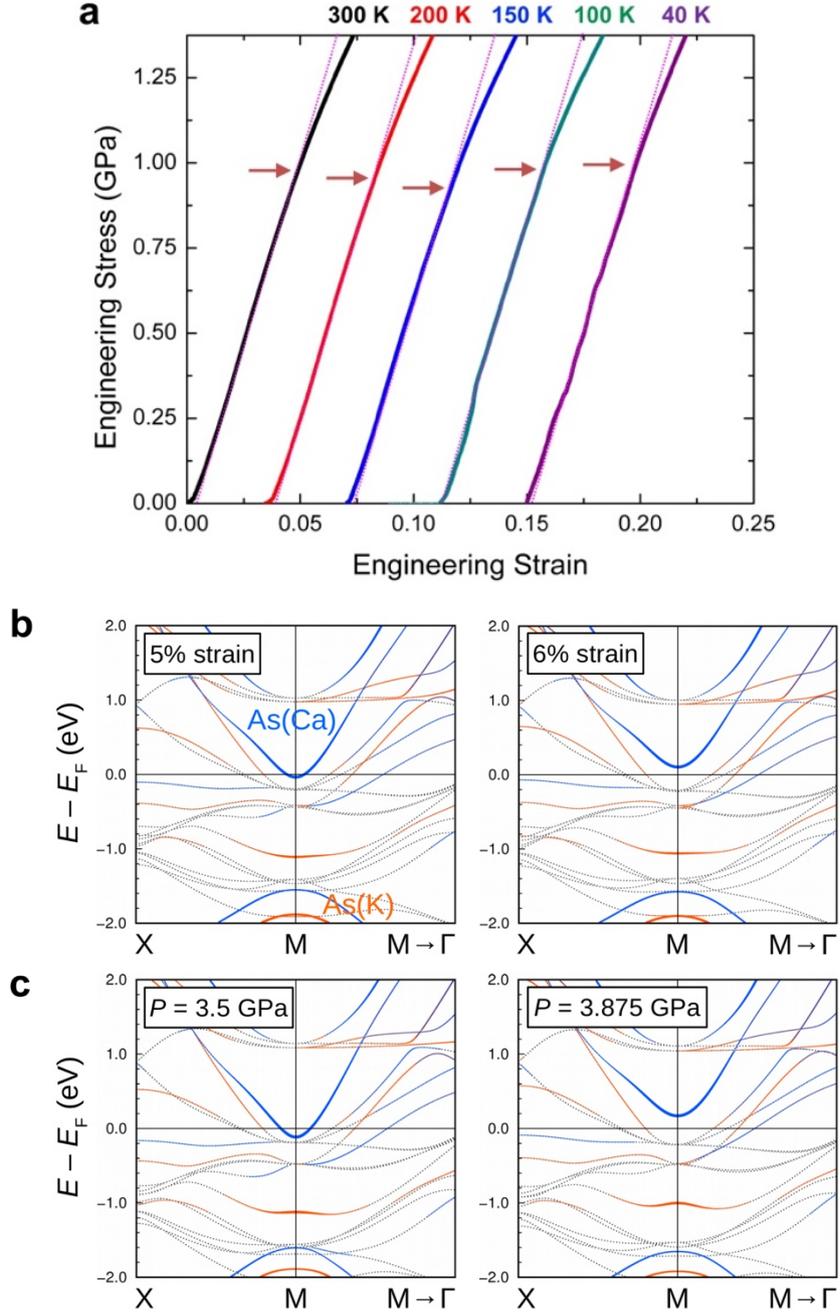

**FIG. 4. Cryogenic nanomechanical test and DFT simulation near the onset of the first hcT transition.** (a) Engineering stress-strain curves of CaKFe$_4$As$_4$ at various cryogenic conditions. The arrow indicates the onset of the first hcT transition; Orbital-resolved non-spin-polarized band structure of CaKFe$_4$As$_4$ under (b) uniaxial and (c) hydrostatic pressure before and after the first hcT (hcT) transition. As $4p_z$ orbitals near the Ca (K) layer are marked by the blue (orange) color. Upon the hcT transition, the antibonding As orbitals shift above the Fermi level. These results show that the change in electronic structure under uniaxial compressive stress (strain) does not differ from that under hydrostatic pressure, implying that the change in electronic properties (here, superconductivity) will be similar under both uniaxial compressive stress and hydrostatic pressure.

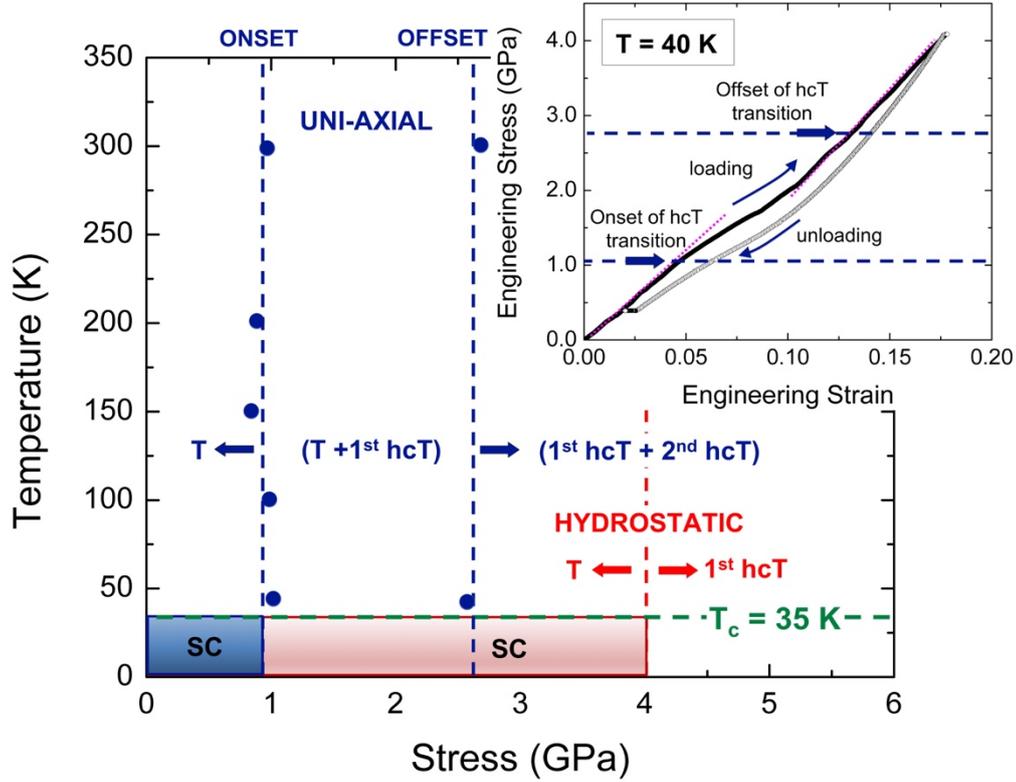

**FIG. 5. Prediction of superconductivity phase diagram in temperature-stress space under uniaxial compression.** At both T= 300 K and 40 K, the hcT transition begins at around 1 GPa of uniaxial stress and complete at around 2.7 GPa. The inset is the loading-unloading curve of $CaKFe_4As_4$ at 40 K, which still shows complete recovery even after ~17% of elastic deformation. The onset stress (~1 GPa) is only 20% of the yield strength (~5 GPa), implying that superconductivity switching would be repeatedly done without failure or even without significant fatigue damage. SC stands for superconductivity. Blue broken lines indicate the onset and offset stresses of the 1st hcT transition under uniaxial compression. Red broken line indicates the onset of 1st hcT transition under hydrostatic compression.[29]

# Supplementary Materials

**Ultrahigh Elastically Compressible and Strain-Engineerable Intermetallic Compounds Under Uniaxial Mechanical Loading**

Gyuho Song,[1] Vladislav Borisov,[2] William R. Meier,[3] Mingyu Xu,[3] Keith J. Dusoe,[1] John T. Sypek,[1] Roser Valentí,[2] Paul C. Canfield,[3] Seok-Woo Lee[1]*

[1]*Department of Materials Science and Engineering & Institute of Materials Science, University of Connecticut, 97 North Eagleville Road, Unit 3136, Storrs CT 06269-3136, USA*

[2]*Institute of Theoretical Physics, Goethe University, Frankfurt am Main, D-60438 Frankfurt am Main, Germany*

[3]*Ames Laboratory & Department of Physics and Astronomy, Iowa State University, Ames IA 50011, USA*

Corresponding author: seok-woo.lee@uconn.edu

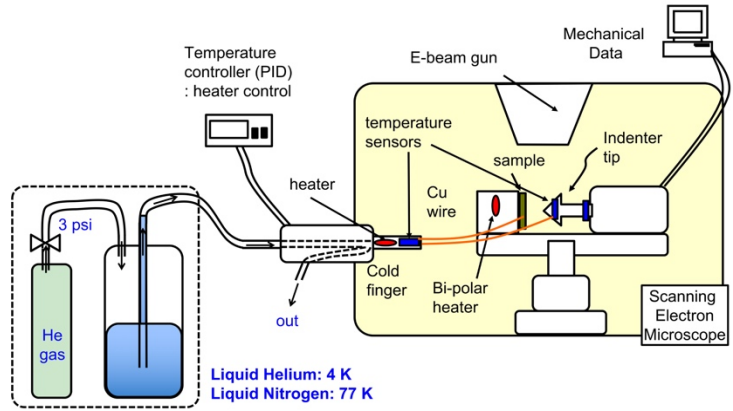
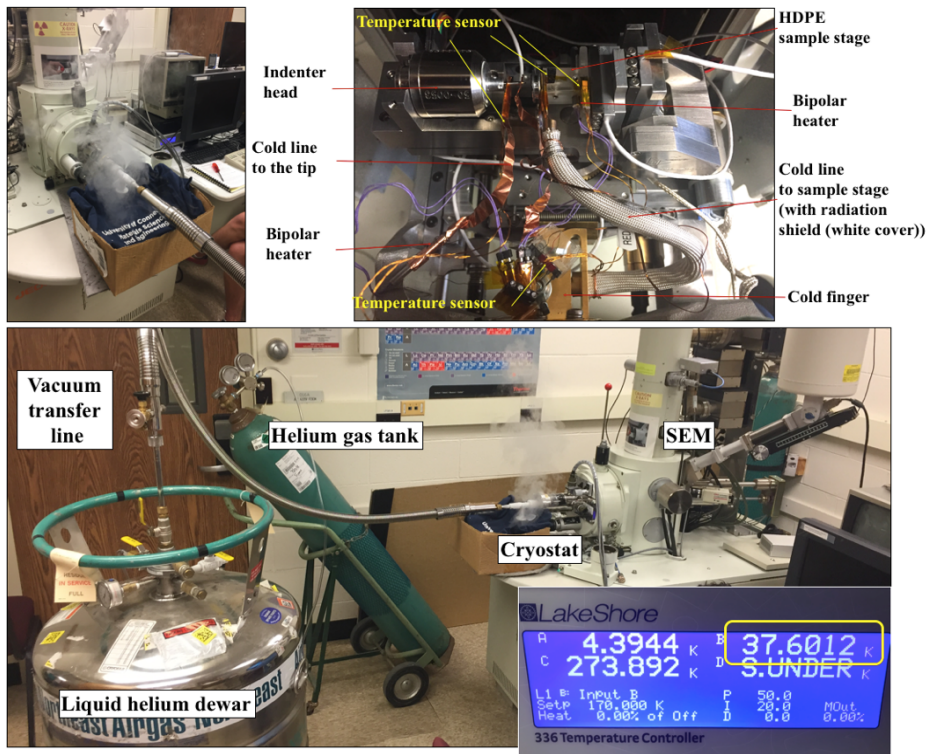

**FIG. S1: State-of-the-art *in-situ* cryogenic nanomechanical testing system with liquid helium cooling capability at University of Connecticut.** He gas is used to control the flow rate of cryogenic liquids. The minimum temperature of the sample stage with liquid helium was about 37 K (See the yellow box in the inset.)

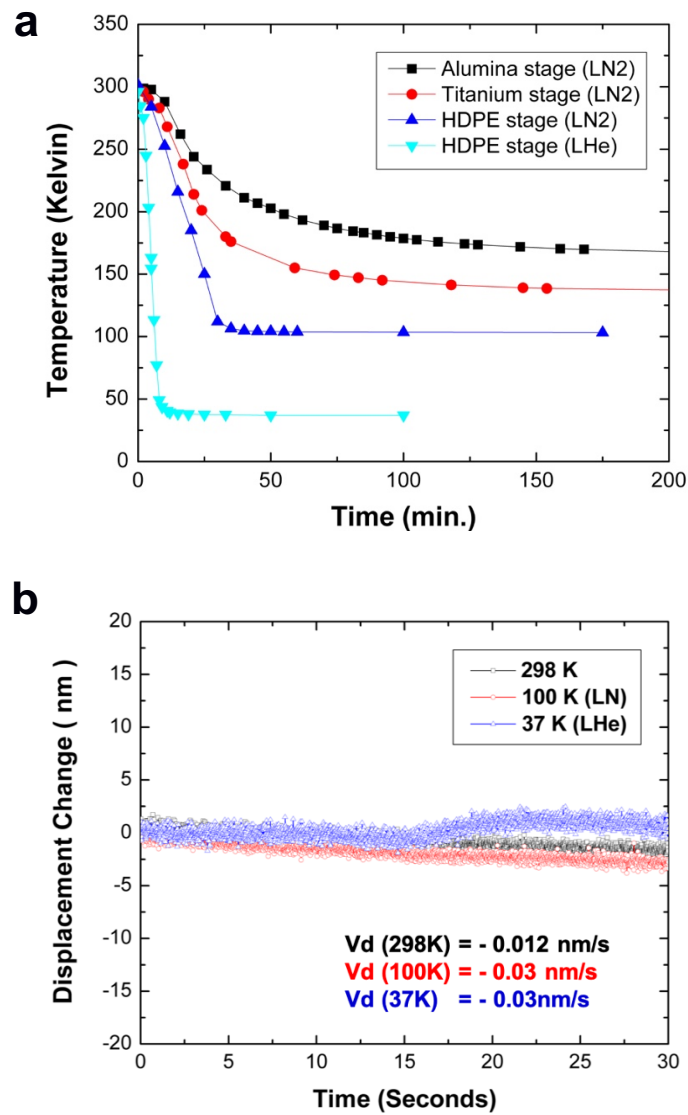

**FIG. S2:** (a) The temperature profile of sample stage with different stage materials. High-Density-Poly-Ethylene (HDPE) shows the best cooling performance due to its lower thermal conductivity (~0.49 Wm$^{-1}$K$^{-1}$) than alumina (~30 Wm$^{-1}$K$^{-1}$) and titanium (~20 Wm$^{-1}$K$^{-1}$); (b) the thermal drift data after cryogenic micromechanical tests. The small thermal drift guarantees the minimal error in displacement.

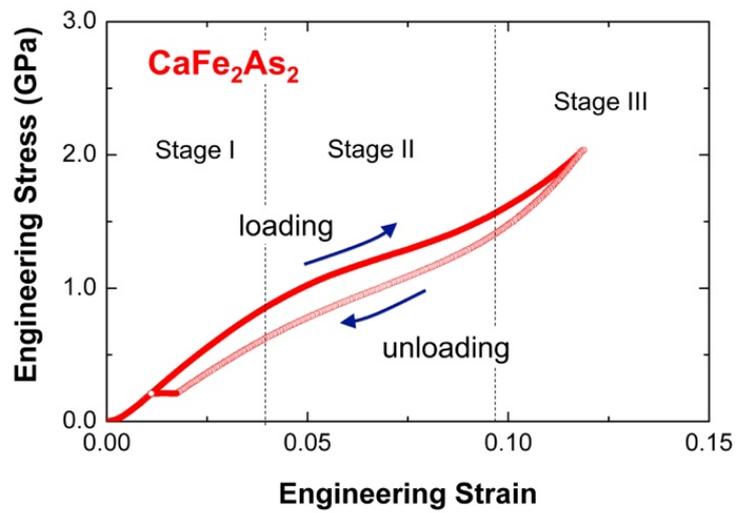

**FIG. S3: Experimental loading-unloading data of $CaFe_2As_2$.** Loading-unloading data of $CaKFe_4As_4$ are also available in Fig. 4(b). Note that both materials exhibit exceptionally large recoverable strains.

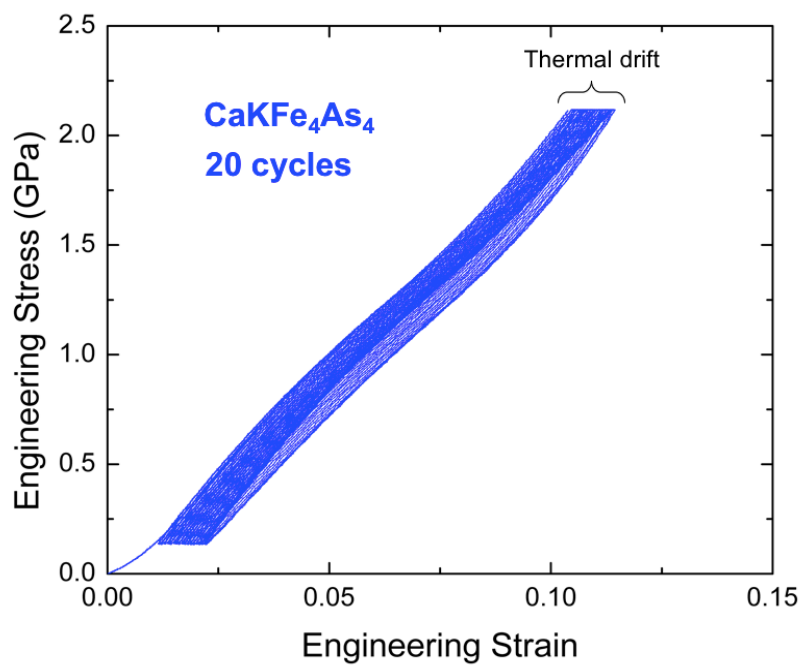

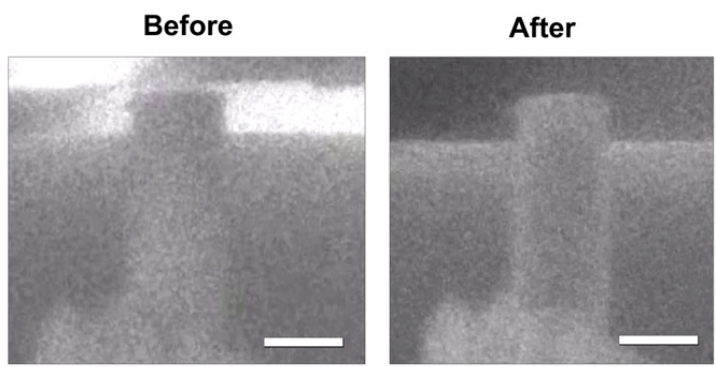

**FIG. S4: Experimental loading-unloading cylic stress-strain curve.** 20 cycles of loading-unloading were applied to CaKFe$_4$As$_4$. The slight shift of stress-strain data over time results from thermal drift, which is one of the common issues associated with long-time nanomechanical testing and is not associated with plastic deformation. Two SEM images below confirm no difference in geometry after 20 loading-unloading cycles (the scale bar, 1.5μm). *In-situ* video of this cyclic deformation is also available as the Supplementary Movie 2.

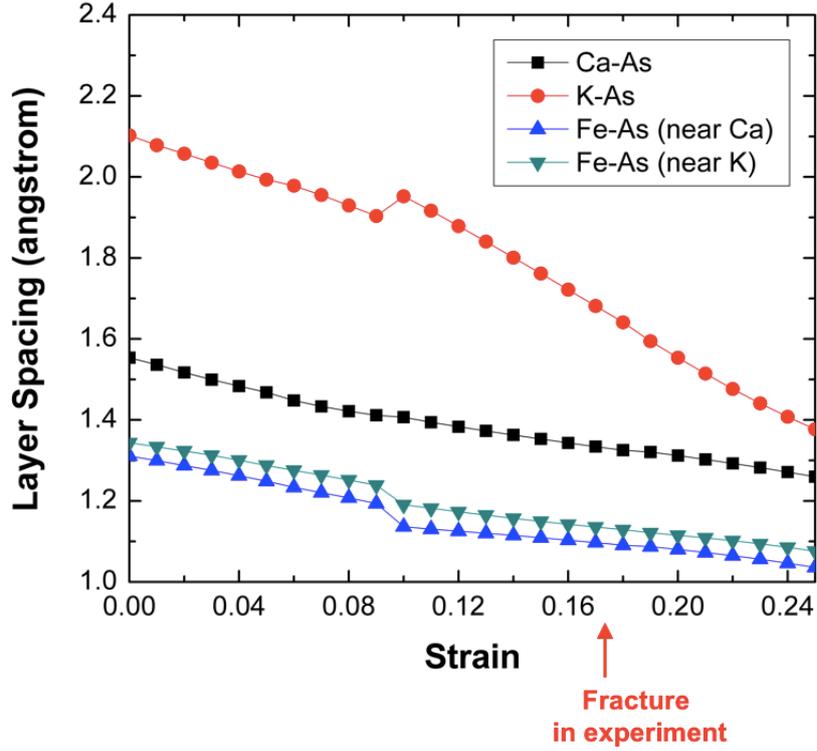

**FIG. S5: DFT data of change in layer spacing of CaKFe$_4$As$_4$ under uniaxial compression.** The layer spacing between K and As layers is decreased the most significantly under uniaxial compression, implying that the region between these two layers are elastically compliant. Based on our calculation, the elastic deformation between As layers around the K atom is responsible for 36% of the total elastic strain. The interlayer distance of As-As layer around a K atom (twice of K-As layer spacing above) is 3.2816Å near the elastic limit. By considering that it is 4.205Å before compression, its change contributes to (4.205Å-3.2816Å)/(12.6206 Å, initial c-length) ≈ 0.073 of strain (~41% of the total elastic limit), which is remarkably high.

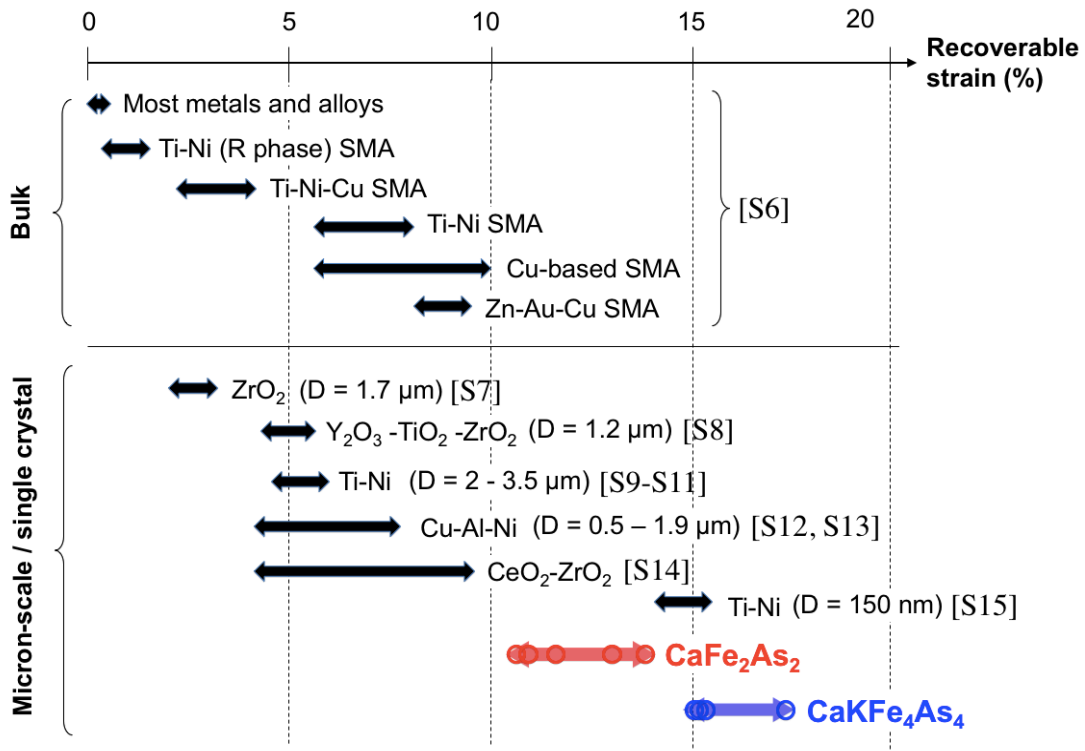

**FIG. S6: Elastic limit of superelastic materials at different length scales with reference numbers** (Fig. 3(a)). The references are available in Supplementary Note 2.

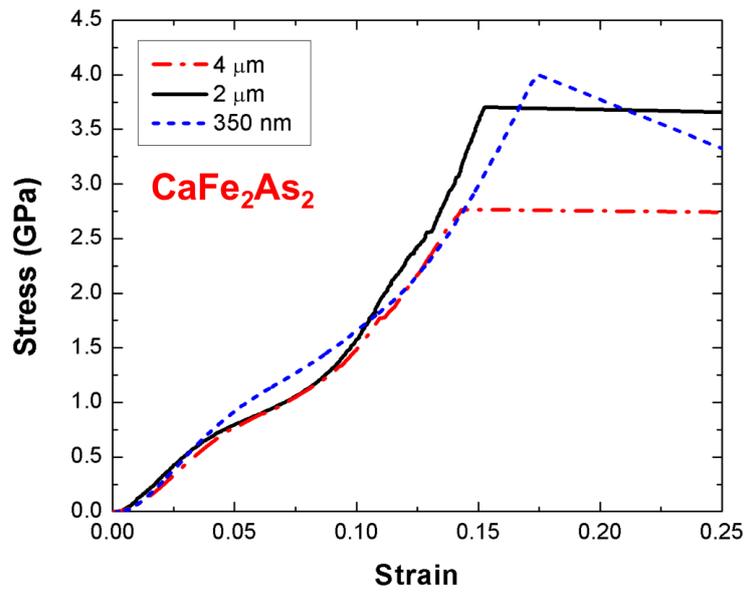

**FIG. S7: Size effect on stress-strain data of CaFe$_2$As$_2$ micropillars with different diameters.** Smaller micropillars exhibit higher yield strength as well as a larger elastic limit. The origin of size effect could be related to the weakest-link mechanism that is related to the statistics of defect distribution; the larger the sample dimension is, the more easily the fracture occurs. FIB damage could also strengthen the surface, leading to larger elastic strain.

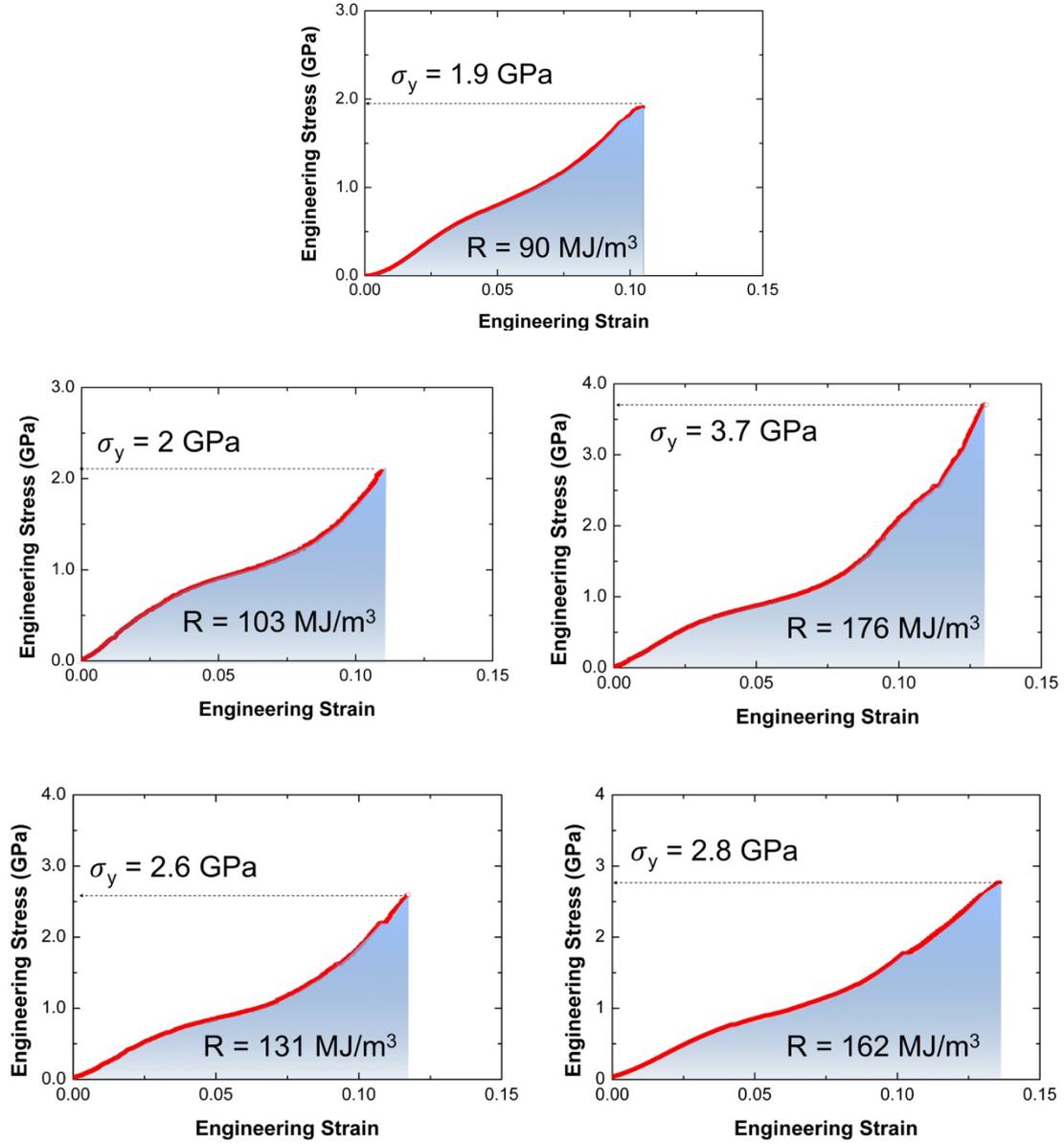

**FIG. S8: Stress-strain data of $CaFe_2As_2$ with the modulus of resilience and yield strength.** Modulus of resilience is obtained by numerical integration of the stress-strain curve between 0 and the elastic limit in strain.

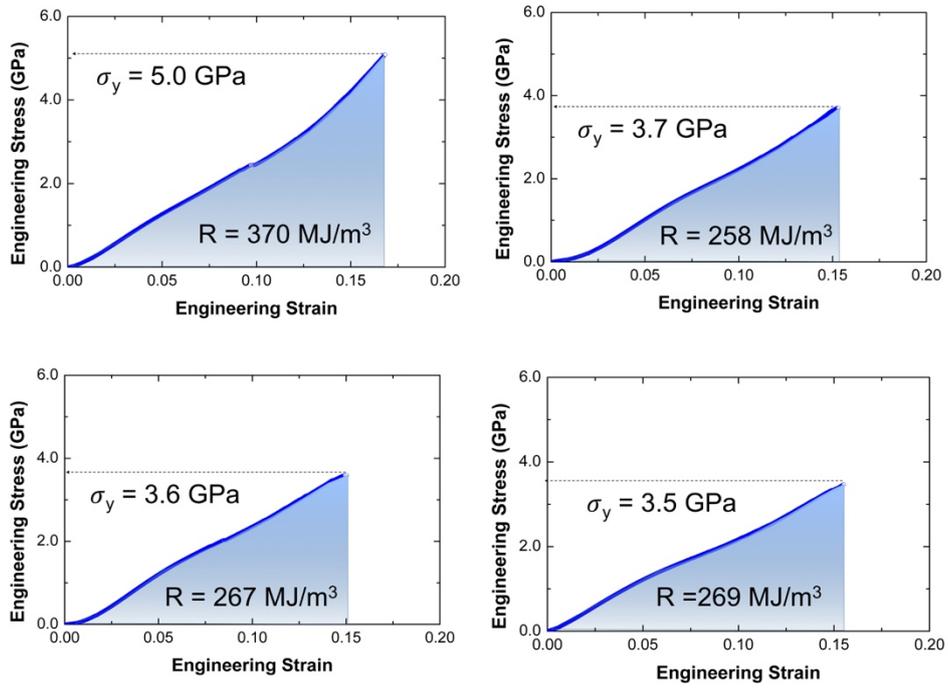

**FIG. S9: Stress-strain data of CaKFe$_4$As$_4$ with the modulus of resilience and yield strength.** Modulus of resilience is obtained by numerical integration of the stress-strain curve between 0 and the elastic limit in strain.

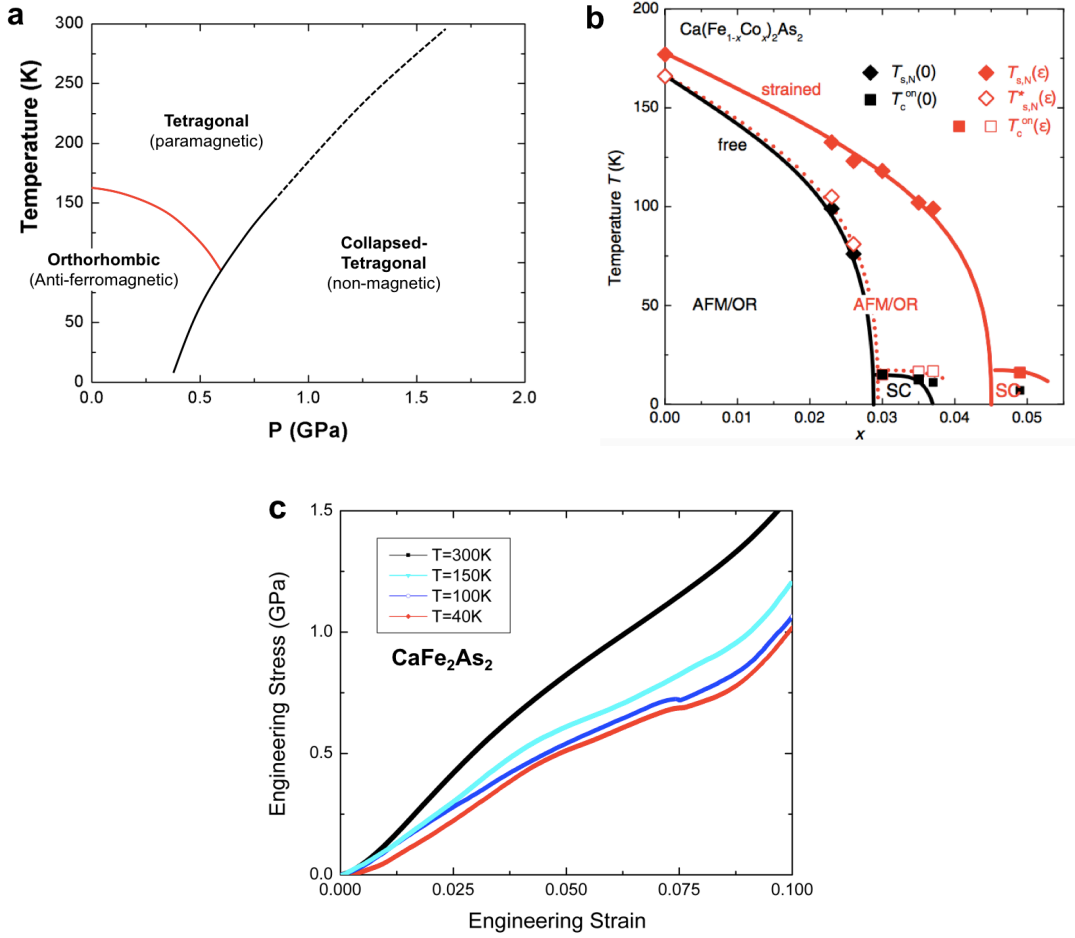

FIG. S10: The temperature sensitivity of $CaFe_2As_2$ and $Ca(Fe_{1-x}Co_x)As_2$. (a) Temperature-pressure phase diagram of $CaFe_2As_2$ under hydrostatic pressure (Reprinted with the permission of J.T. Sypek *et al.*[29], copyright 2017, Springer Nature); (b) Temperature-composition phase diagram of $Ca(Fe_{1-x}Co_x)As_2$ (Reprinted with the permission of A.E. Böhmer *et al.*[42], copyright 2017, American Physics Society); (c) (**This study**) temperature-dependent engineering stress-strain curve of $CaFe_2As_2$ under uniaxial compression. In contrast to the temperature insensitivity of stress-strain data of $CaKFe_4As_4$ (Fig. 4(a)), stress-strain data of $CaFe_2As_2$ is strongly dependent on temperature. As the temperature decreases, the cT transition occurs at a lower stress. This result is consistent with the result of the hydrostatic pressure experiment in Ref. 35.

**Supplementary Note 1: Development of *in-situ* nanomechanical testing system**

An *in-situ* cryogenic nanomechanical testing system has been developed to investigate the effects of temperature on superelasticity in $CaKFe_4As_4$. Temperatures ranging from room temperature down to 40 K were reached using liquid He (LHe) and a customized Janis cryogenic cold finger system (Janis Research Company, MA, USA) combined with the *in-situ* micropillar compression system (Fig. S1). The cold finger itself reaches ~4 K, the minimum temperature of LHe, but there is always cold loss due to heat transfer from the sample stage and radiation from the chamber wall. In order to mitigate this and reach a minimum temperature of 40K, multiple iterations of the stage design and connections needed to be done. The overall cryogenic system design is similar to that of the alumina and titanium stage.[S45-S47] Note that the cooling capability of the sample stage is strongly dependent of the sample stage material. In this work, we developed a high-density polyethylene (HDPE) stage due to its thermal conductivity (~0.49 $Wm^{-1}K^{-1}$) being much lower than alumina (~30 $Wm^{-1}K^{-1}$) and titanium (~20 $Wm^{-1}K^{-1}$). In previous works, we confirmed that it takes more than 10 hours to achieve 130 K with alumina and titanium stages. With our new HDPE sample stage, however, we were able to achieve 40 K only within one hour because HDPE provided the least amount of thermal conduction for the system (Fig. S2). Thus, the HDPE sample stage has a great cooling performance and is stiff enough for reliable mechanical testing.

Micropillar compression testing was performed at 40 K, 100 K, 150 K and room temperature. During the cryogenic test, we carefully managed the thermal drift by equilibrating the tip temperature with the sample temperature. This approach is the same with the technique suggested by the literatures of high temperature nanoindentation and compression testing of shape-memory alloys and ceramics.[S48,S49] It was demonstrated that the temperature equilibration of the tip through mechanical contact could be used up to 500 °C for the high temperature nanoindentation. For cryogenic testing, the maximum possible temperature difference from room temperature is only 300 K and is about 260 K (= 300 K – 40 K) in this study. Thus, it is expected that the tip contact cooling would work better in the case of cryogenic test. For our testing purposes, the flat punch tip holds contact with the cooled sample surface for at least, if not more than, 30 minutes to equilibrate the tip and sample temperatures. By monitoring the change in displacement under a constant load condition, we were able to confirm that the temperature profile reaches steady state. Then, the tip is immediately translated to a micropillar, and the compression test is conducted quickly. In using this method, post testing data shows that the most thermal drift measured is always below 1 nm/s, which is small enough to assume no significant change in temperature during mechanical testing.

**Reference of Supplementary Note 1**

**Supplementary Note 2: References of Figure 3 in the main text**

The following is the references of materials in Fig. 3 in the main text. Reference number is included in Fig. S4.

[S6]T. Omori, R. Kainuma, Nature 502, 42 (2013).

[S7]Z. Du, X. M. Zeng, Q. Liu, A. Lai, S. Amini, A. Miserez, C. A. Schuh, C. L. Gan, Scripta Mater. **101**, 40 (2015).

[S8]X. M. Zeng, A. Lai, C. L. Gan, C. A. Schuh, Acta Mater. **116**, 124 (2016).

[S9]C. P. Frick, B. G. Clark, A. S. Schneider, R. Maaß, S. Van Petegem, H. Van Swygenhoven, Scripta Mater. **62**, 492 (2010).

[S10]C. P. Frick, S. Orso, E. Arzt. Acta Mater. **55**, 3845 (2007).

[S11]J. Pfetzing-Micklich, R. Ghisleni, T. Simon, C. Somsen, J. Michler, G. Eggeler, Mater Sci Eng: A **538**, 265 (2012).

[S12]J. San Juan, M. L. Nó, C. A. Schuh, Acta Mater. **60**, 4093 (2012).

[S13]J. San Juan, M. L. Nó, C. A. Schuh, Adv Mater. **20**, 272 (2008).

[S14]Lai, Z. Du, C. L. Gan, C. A. Schuh, Science **341**,1505 (2013).

[S15]J. Ye, R. K. Mishra, A. R. Pelton, A. M. Minor, Acta Mater. **58**, 490 (2010).

**Supplementary Note 3: Modulus of Resilience of various nano-sized materials**

| Material | | | σ (GPa) | E (GPa) | R (MJ/m$^3$) | Reference |
|---|---|---|---|---|---|---|
| Metallic Nanopillar | BCC | V | 2.45 | 128 | 23.5 | [S16] |
| | | Nb | 0.76 | 105 | 2.7 | [S17] |
| | | Mo | 3.96 | 329 | 23.8 | [S18] |
| | | Nb | 1.15 | 105 | 6.3 | [S18] |
| | | Ta | 2.77 | 186 | 20.6 | [S18] |
| | | W | 2.85 | 411 | 9.9 | [S18] |
| | | Ta | 1.49 | 186 | 6.0 | [S16] |
| | HCP | Ti | 3.27 | 116 | 46.0 | [S19] |
| | | Ti | 3.28 | 116 | 46.3 | [S19] |
| | | Mg | 0.58 | 45 | 3.7 | [S20] |
| | | Mg | 0.36 | 45 | 1.4 | [S21] |
| | FCC | Cu | 2.77 | 120 | 32.0 | [S22] |
| | | Cu | 2.82 | 120 | 33.2 | [S23] |
| | | Au | 1.06 | 79 | 7.1 | [S24] |
| | | Au | 1.36 | 79 | 11.7 | [S25] |
| Nanowire | | Si | 18.5 | 161 | 1064 | [S26] |
| | | Si | 20 | 125 | 1600 | [S27] |
| | | Si | 12.2 | 188 | 397 | [S28] |
| | | Ge | 18 | 106 | 1530 | [S29] |
| | | Ge | 15 | 200 | 563 | [S30] |
| | | ZnO | 9.5 | 153 | 295 | [S31] |
| | | ZnO | 12.1 | 173 | 424 | [S32] |
| | | ZnS | 0.37 | 5 | 15 | [S33] |
| | | GaAs | 5.4 | 77 | 189 | [S34] |
| | | InAs | 5 | 50 | 250 | [S35] |
| | | GaN | 3.1 | 124 | 39 | [S36] |
| | | WS$_2$ | 16 | 114 | 1120 | [S37] |
| | | VO$_2$ | 5.2 | 137 | 99 | [S38] |
| | | α-Al$_2$O$_3$ | 48.8 | 460 | 2586 | [S39] |
| | | SiC | 35 | 500 | 1225 | [S40] |
| | | SiC | 53.4 | 534 | 2670 | [S41] |
| | | Au | 8 | 80 | 400 | [S42] |
| | | Au | 9.8 | 50 | 960 | [S43] |
| | | Cu | 5.8 | 81 | 209 | [S44] |
| | | Cu nanotwin | 2.12 | 42 | 53 | [S45] |
| | | Ni | 5 | 14 | 865 | [S46] |
| | | Nb | 1.8 | 45 | 36 | [S47] |
| | | Ag | 4.8 | 120 | 96 | [S48] |
| | | Co | 2.04 | 95 | 22 | [S49] |

**Supplementary Movie 1. Uniaxial deformation of CaKFe$_4$As$_4$ until fracture occurs.** Total fracture strain is about 17% (See also the snapshot in Fig. 2(a))

**Supplementary Movie 2. Cyclic compression test of CaKFe$_4$As$_4$ (20 cycles).** The cyclic compression strain is about 11%, and this large deformation is recoverable. The stress-strain data and video snapshots are available in Fig. S2.